\documentclass[sigconf,nonacm]{acmart}

\AtBeginDocument{%
  \providecommand\BibTeX{{%
    \normalfont B\kern-0.5em{\scshape i\kern-0.25em b}\kern-0.8em\TeX}}}

\newcommand\ruleeq{\mathrel{\mathord{:}-}}

\theoremstyle{definition}

\newtheorem{example}{Example}[section]

\begin{document}

\title{Toward Co-existing Database Schemas based on Bidirectional Transformation}



\author{Jumpei Tanaka}
\email{jumpeitanaka@nii.ac.jp}
\affiliation{
    \institution{The Graduate University for Advanced Studies, SOKENDAI, Japan}
}

\author{Van-Dang Tran}
\email{dangtv@nii.ac.jp}
\affiliation{
    \institution{National Institute of Informatics, Japan}
    \institution{The Graduate University for Advanced Studies, SOKENDAI, Japan}
}

\author{Hiroyuki Kato}
\email{kato@nii.ac.jp}
\affiliation{
    \institution{National Institute of Informatics, Japan}
}

\author{Zhenjiang Hu}
\email{huzj@pku.edu.cn}
\affiliation{
    \institution{Peking University, China}
    \institution{National Institute of Informatics, Japan}
}

\begin{abstract}
 According to strong demands for rapid and reliable software delivery, co-existing database schema versions with multiple application versions are reality to contribute them.
Current database management systems do not support co-existing schema versions in one database.
Although a design of co-existing schema based on updatable view tables was previously proposed, its flexibility is limited due to pre-defined several restrictions to achieve data synchronization among schemas and handling independent unsynchronized data in each schema.
In this preliminary report, we present a new approach for co-existing schemas based on bidirectional transformation.
We explain the required properties to realize co-existing schemas, {\em bidirectionality} and {\em totality}.
We show that the co-existing schemas can be implemented systematically by applying putback-based bidirectional transformation to satisfy both the bidirectionality and the totality. While the bidirectionality can be satisfied by applying bidirectional transformation, to satisfy the totality, extra functions need to be introduced. How to derive these extra functions is presented.
\end{abstract}






\maketitle

\section{Introduction}
Current database management systems do not support co-existing schema versions in one database.
Along with a constant flux of business requirements and demand for rapid, reliable and evolutionary software development, recent efforts of continuous delivery, e.g. DevOps and Agile, have developed version control technique to maintain multiple versions of application \cite{Humble}.
In evolutionary steps co-existing multiple application versions makes evolution faster, safer and more reliable.
It achieves to deliver new features by a new version while a previous version runs steadily in parallel. A new version can be checked by several users as pilot until it is fully warmed up while other users steadily enjoy existing features provided by a previous application version\cite{Fitzgerald}.
GIT, SVN and so on are strong tools for application version control.

However, it is hard for a single database to run multiple schema versions concurrently even if co-existing each application version expects to have the corresponding schema version which works like a full-fledged database.
Normally a database runs only one schema version with its physical data.
Having a new schema version a database forces to migrate physical data from a previous schema version.
This work is executed by a manually written roll-forward script which maps structures between schemas\cite{Humble}.
This is expensive and error-prone work.
Furthermore, to keep a previous schema version up and running, roll-back data migration by another manually written script is required.
Due to these painful and error-prone works, co-existing schemas is not pragmatic today. Software development with application and database has limitation of its productivity.
Co-existing schema versions in a single database is reality to notably contribute productivity, efficiency and reliability of evolutionary software development.
Note that while schema evolution does not require a previous schema after evolution\cite{Jagadish}, co-existing schemas requires both a previous schema and an evolved new schema to behave like a full-fledged database on each schema concurrently.

To achieve co-existing schemas on relational database, Multi-Schema-Version Database Management Systems (MSVDB) was proposed by K. Herrmann, et al. \cite{Herrmann1, Herrmann2}.
This shows a necessary work for keeping multiple schema versions up and running concurrently.
It is required {\bf to accept any update on each schema, to synchronize updated data among schemas for necessity and to keep update of unsynchronized data in each schema} over one physical data.

\begin{figure}[t]
    \includegraphics[scale=0.26]{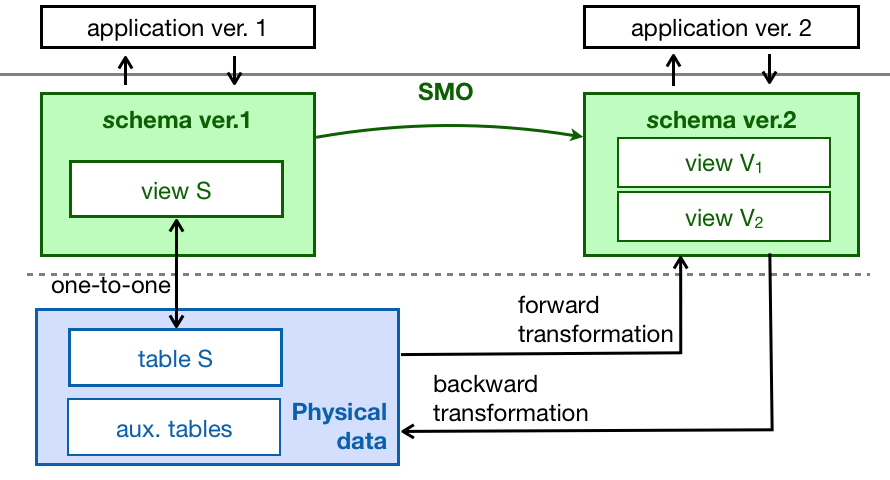}
    \caption{MSVDB}
    \label{fig:MSVDB}
\end{figure}

To achieve these features, MSVDB constructs tables in each schema by updatable views.
Figure \ref{fig:MSVDB} shows its structure.
The schema ver.1 as original has the table ${\rm S}$ as an updatable view which has one-to-one mapping from physical data of table ${\rm S}$.
The table ${\rm V_1}$ and ${\rm V_2}$ in the schema ver.2 are evolved from the table ${\rm S}$ in schema ver.1.
These tables are also updatable views which is computed by forward transformation from physical data of table ${\rm S}$ and auxiliary tables.
Evolution is not necessarily information-preserving against an original table.
For example, adding a new column and separately maintaining unsynchronized data  require a setup to preserve data separately from an original table otherwise the data is lost.
Auxiliary tables preserve this auxiliary information.
Thus backward transformation reflects update on table ${\rm V_1}$ and ${\rm V_2}$ in schema ver.2 to physical data of table ${\rm S}$ and auxiliary tables.
Schema evolution is given by primitive operators called SMO (Schema Modification Operations), e.g. adding a new column or splitting row.
SMO creates new tables for a new schema and generates rules of forward and backward transformation.

This design causes difficulty more than well-known {\it view update problem} which treats synchronization between an updatable view and physical data \cite{Bancilhon}.
In the previous work of MSVDB, the author narrowed down this problem into a limited number of primitive operators, SMOs, without treating arbitrary evolved schemas.
By partially using Bidirectional Transformation technique to check well-behaveness of data synchronization\cite{Hofmann}, the author carefully and heuristically designs forward and backward transformation of each SMO due to lack of a generalized method to uniquely derive rules for the both synchronized and unsynchronized data.
Unfortunately, the some ad-hoc behavioral restrictions, e.g.
primary key as mandatory and complicatedly prepared auxiliary tables, makes user's flexibility limited.
A user is forced to take care these behaviors and restrictions in SMO and design an intended schema by applying several SMOs one-by-one.

In this preliminary report, we present a new approach for co-existing schemas by fully applying Bidirectional Transformation (BX).
BX is a matured technique for data synchronization \cite{Foster}.
By introducing a new approach to handle unsynchronized data with BX technique, a user is able to directly and arbitrarily design co-existing schemas.
More specifically, a user writes backward transformation for synchronized data then rules of backward transformation and forward transformation for the both synchronized and unsynchronized data are uniquely derived so that
co-existing schemas is achieved.
We contribute to improve user's flexibility to have co-existing schemas.

\section{Feature of Co-existing Schemas}
In this section, the required feature of co-existing schemas is shown based on the architecture of MSVDB in Figure \ref{fig:MSVDB}.
Then it is formatted to property of forward and backward transformations.
First, an intuitive example is introduced.

\begin{figure}[t]
    \includegraphics[scale=0.45]{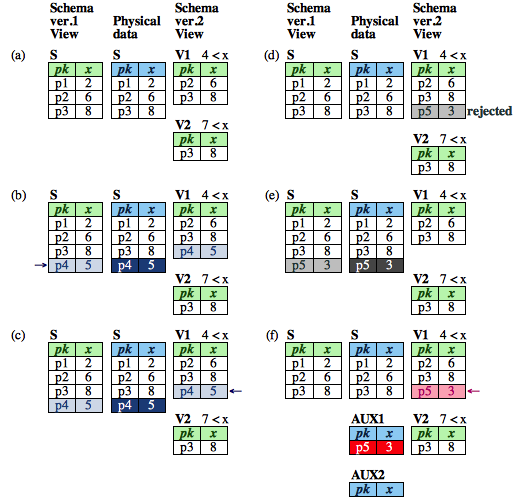}
    \caption{example of co-existing schemas}
    \label{fig:sample}
\end{figure}

\begin{example}
    Let $\mathrm{S}(pk, x)$ be a table which has attribution $pk$ as primary key and contains physical data.
    Schema ver.1 is consisted by view $\mathrm{S}(pk, x)$ as one-to-one mapping from physical data of $\mathrm{S}(pk, x)$.
    Schema ver.2 is consisted by view $\mathrm{V_1}(pk, x)$ and $\mathrm{V_2}(pk, x)$.
    They are defined as selection of $\mathrm{S}(pk, x)$: $\mathrm{V_1} = \sigma_{4 < x} (\mathrm{S})$ and $\mathrm{V_2} = \sigma_{7 < x} (\mathrm{S})$.
    Figure \ref{fig:sample} (a) shows initial computation of views in both schemas.
\end{example}

Figure \ref{fig:sample} (b) shows insertion of synchronized data on view ${\rm S}$ of schema ver.1.
When $(p4, 5)$ is inserted into view ${\rm S}$, it is inserted to table ${\rm S}$ as physical data due to one-to-one mapping.
Then instance of views in schema ver.2 is computed from the updated physical data.
${\rm V_1}$ shows $(p4, 5)$ because $4 < x$ is satisfied and ${\rm V_2}$ does not show it because $7 < x$ is not satisfied.
Similarly Figure \ref{fig:sample} (c) shows insertion of synchronized data on view ${\rm V_1}$ of schema ver.2.
When $(p4, 5)$ which satisfies $4 < x$ is inserted into view ${\rm V_1}$, this insertion can be inserted to table ${\rm S}$ as physical data because it can survive in one round trip to recompute instance of view ${\rm V_1}$ from the updated physical data.
View ${\rm S}$ of schema ver.1 also shows $(p4, 5)$ as one-to-one mapping.
This feature of synchronization is formatted as {\bf bidirectionality}.
This is a property of a pair of forward and backward transformation to synchronize update between source as physical data and view.

Figure \ref{fig:sample} (d) shows insertion of unsynchronized data on view ${\rm V_1}$ of schema ver.2.
This operation intends that schema ver.2 manages unsynchronized data independently from schema ver.1.
When $(p5, 3)$ which does not satisfy $4 < x$ is inserted into view ${\rm V_1}$, this insertion is rejected by backward transformation which satisfies bidirectionality.
As Figure \ref{fig:sample} (e) shows, even if $(p5, 3)$ is inserted into table $\mathrm{S}$ as physical data, recomputed view $\mathrm{V_1}$ does not show it anymore because it does not satisfy $4 < x$.
Thus backward transformation to satisfy bidirectionality does not accept insertion of $(p5, 3)$.
Additional design is expected to handle unsynchronized data.

Figure \ref{fig:sample} (e) shows a setup to handle unsynchronized data.
Auxiliary tables $\mathrm{AUX_1}(pk, x)$ and $\mathrm{AUX_2}(pk, x)$ are prepared to store insertion of unsynchronized data on ${\rm V_1}$ and ${\rm V_2}$ respectively.
Backward transformation is additionally designed to reflect an insertion into auxiliary tables when it does not satisfy condition of bidirectionality for synchronized data.
For example, when $(p5, 3)$ is inserted into $\mathrm{V_1}$, it is inserted into table $\mathrm{AUX1}$ as physical data.
By designing forward transformation as union of $\sigma_{4 < x} (\mathrm{S})$ and $\mathrm{AUX1}$, the recomputed instance of ${\rm V_1}$ contains $(p5, 3)$.
View ${\rm S}$ does not show $(p5, 3)$ because it is computed only from physical data of ${\rm S}$.
This feature is formatted as {\bf totality} of backward transformation.
This is a property of backward transformation to accept any insertion on view and reflect to source as physical data regardless synchronized data or unsynchronized data.

This example shows that co-existing schemas is achievable by designing source tables for physical data and a pair of forward and backward transformation.
In MSVDB, they are heuristically designed by the author.
In this preliminary work, we propose a generalized method by applying bidirectional transformation (BX) technique.

\section{Proposed approach}
In this section, the proposed approach is presented to implement co-existing schemas systematically.
To this end, putback-based BX is applied to satisfy both bidirectionality and totality mentioned in the previous section.
Bidirectionality can be satisfied by just applying a well-behaved BX as you will be seen in Section~\ref{sec:bx}.
On the other hand, to satisfy the totality extra functions need to be introduced.
How to automatically derive these functions is described in Section~\ref{sec:derivetotalfunctions}.
The whole steps to implement co-existing schemas are described in Section~\ref{sec:whole} followed by an example in Section~\ref{sec:ex}.

\subsection{Bidirectional Transformation (BX)}
\label{sec:bx}
Oriented from {\it view update problem} \cite{Bancilhon}, a bidirectional transformation (BX) is a matured technique to synchronize between two data \cite{Foster}.
BX consists of a pair of forward and backward transformation between a set of source data $S$ and a set of view data $V$.
The forward transformation $get(s) = v$ accepts a source $s \in S$ and produces a view $v \in V$.
The backward transformation $put(s, v')= s'$ accepts an original source $s$ and an updated view $v' \in V$, and produces an updated source $s' \in S$.
As contrasted with co-existing schemas, $get$ function corresponds to forward transformation to compute views from physical data and $put$ function corresponds to backward transformation to propagate update on view to physical data.
To satisfy bidirectionality, a pair of transformations should be well-behaved in the sense they satisfy following round-tripping laws.

$$
\begin{array}{ll}
     put (s, get(s)) = s & (GetPut) \\
     get(put (s, v')) = v' & (PutGet)
\end{array}
$$

The {\it GetPut} property requires that no change on the view should result in no change on the source, while the {\it PutGet} property demands change of the view to be translated to the source and the updated view to be the one computed from the updated source.

It is also known that defining $get$ first causes ambiguity to derive well-behaved pair of $put$ \cite{Fischer}.
To avoid this ambiguity, putback-based BX has been well studied \cite{Ko, Van-dang2}.
It defines $put$ first then uniquely derive well-behaved pair of $get$.
Due to usability to uniquely derive a pair of transformations, putback-based BX recently demonstrated the application of database to synchronize data among diversified source tables\cite{Asano, Ishihara}.

In our proposed approach for co-existing schemas, putback-based BX is utilized so that a user can arbitrary define a new schema and backward transformation then a suitable pair of forward and backward transformation is uniquely derived.
Describing backward transformation means a user writes a rule to embed an evolved new table into original tables by having update propagation rule between them.

\subsection{Satisfying Totality of Backward Transformation}
\label{sec:derivetotalfunctions}
To apply BX to co-existing schema, an extra function is introduced for totality of backward transformation.

\begin{figure}[t]
    \includegraphics[scale=0.50]{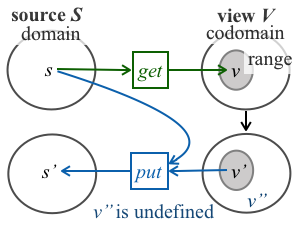}
    \caption{range and codomain}
    \label{fig:range}
\end{figure}

Example 2.1 shows forward transformation as selection which discards information of source.
Normally forward transformation $get$ discards information when it produces the view $v$ \cite{Matsuda}.
Thus range of $get$, $r\!ange (get)$, is subset of its codomain $V$ (Figure \ref{fig:range}).
To satisfy round-tripping law, acceptable view update should be in range of $get$.
Otherwise, even if update out of $r\!ange(get)$ is reflected to the source, this update is discarded when $get$ recomputes the view from the updated source.
$put$ is defined in $r\!ange (get)$ as subset of $V$.
This is a partial function.
$$
\begin{array}{ll}
put ::  S \times r\!ange (get) \rightarrow S \\
r\!ange (get) \subset V
\end{array}
$$

Therefore if we construct backward transformation of co-existing schema only by a partial function $put$, totality of backward transformation is not satisfied.
We introduce a method to make backward transformation total based on $put$ and $get$ of BX.


To explicitly treat view update, an updated view is decomposed into three relations: a set of tuples in a current view ($\rm{V^{cur}}$), a set of inserted tuples ($+{\rm V}$) and a set of deleted tuples ($-{\rm V}$).
An updated view ${\rm V}$ is constructed by removing all tuples of deletion ($-{\rm V}$) from a current view $\rm{V^{cur}}$ and adding all tuples of insertion ($+{\rm V}$).
This operation is expressed by the following Datalog rules, where $\vec{x}$ denotes a tuple of variables.

$$
\left\{
\begin{array}{lrlr}
     {\rm V}(\vec{x}) & \ruleeq & {\rm V^{cur}}(\vec{x}), \; \neg -{\rm V}(\vec{x}). & {\qquad} (1) \\
     {\rm V}(\vec{x}) & \ruleeq & +{\rm V}(\vec{x}). & {\qquad} (2)
\end{array}
\right.
$$

$put$ is defined for update of synchronized data to compute insertion and deletion of source ($+{\rm S}$ and $-{\rm S}$) when given insertion and deletion of view ($+{\rm V}$ and $-{\rm V}$) are in $r\!ange(get)$.
Conversely undefined update in $put$ is update of unsynchronized data.
To make backward transformation total and to handle unsynchronized data, undefined case of $put$ is formalized as an extra function $unde\!f$.
$unde\!f$ computes insertion and deletion of view ($+{\rm V^{ud}}$ and $-{\rm V^{ud}}$) which do not compute $+{\rm S}$ and $-{\rm S}$ by $put$.
This is defined by the following Datalog rules.

$$
\left\{
\begin{array}{lrlr}
    {\pm}{\rm S}(\vec{x}) & \mathrel{\mathord{:}-} & +{\rm S}(\vec{x}). & {\qquad} (3)\\
    {\pm}{\rm S}(\vec{x}) & \mathrel{\mathord{:}-} & -{\rm S}(\vec{x}). & {\qquad} (4) \\
	+{\rm V^{ud}}(\vec{x}) & \mathrel{\mathord{:}-} & +{\rm V}(\vec{x}), \; \neg {\pm}{\rm S}(\vec{x}). & {\qquad} (5) \\
	-{\rm V^{ud}}(\vec{x}) & \mathrel{\mathord{:}-} & -{\rm V}(\vec{x}), \; \neg {\pm}{\rm S}(\vec{x}). & {\qquad} (6)
\end{array}
\right.
$$

By constructing backward transformation with $put$ and ${unde\!f}$, it satisfies totality.
As contrasted with co-existing schemas, ${\rm V^{ud}}(\vec{x})$ becomes the auxiliary table of physical data to store update of unsynchronized data on view in a new schema.

\subsection{Steps to Derive Rules for Co-existing Schemas}
\label{sec:whole}
As input, a user arbitrarily writes a rule of $putdelta(s, v)$ by Datalog to produce $(+s, -s)$.
$putdelta$ is utilized instead of $put$ to clearly describe delta (insertion or deletion) of $S$ and directly corresponds to trigger of SQL for implementation purpose.
As output, two rules written by Datalog are derived, $unde\!f(+v, -v)$ as an extra function of backward transformation to produce $v^{ud} \in V^{ud}$ and $get'(s, v^{ud})$ as a function of forward transformation to produce view $v \in V$.
A pair of forward transformation as $get'$ and backward transformation as $putdelta$ and $unde\!f$ satisfies bidirectionality and totality.

A method to derive output is as following steps.
To derive $get$ from $putdelta$, the putback-based BX with Datalog, BIRDS, is utilized \cite{Van-dang1}.

\vspace{\baselineskip}

\begin{itemize}
\item {\bf step 1}: \
$putdelta(s, v)$ is input into the putback-based BX.
It derives $get(s) = v$ which becomes well-behaved BX and satisfies bidirectionality.

\item {\bf step 2}: \ $putdelta(s, v)$ is deformed as $putdelta'(s, v^{cur}, +v, -v)$ by substituting (1) and (2).

\item {\bf step 3}:\  A rule of $unde\!f(+v, -v) = v^{ud} $ is derived from definition of $unde\!f$ ((3) - (6)), $putdelta'(s, v^{cur}, +v, -v)$ and $get(s)$. In a case of selection, rules containing equation or non-equation against update of view are utilized.

\item {\bf step 4}: \ $putdelta$ and derived $unde\!f$ as total backward transformation are put into the putback-based BX.
It derives $get'(s, v^{ud}) = v$ of forward transformation which satisfies bidirectionality.
\end{itemize}

\vspace{\baselineskip}

Finally, the putback-based BX automatically translates derived Datalog rules of forward and backward transformation to SQL.

\subsection{Example}
\label{sec:ex}
From Example 2.1 selection between $\rm{S}$ and $\rm{V_1}$ is used to demonstrate the proposed method.

\vspace{\baselineskip}

\begin{description}
  \item[Input]
    \begin{itemize}
        \item []
        \item  Source: ${\rm S}(x)$
        \item  View:   ${\rm V_1}(x)$
        \item  $putdelta$
        $$
        \begin{array}{lrl}
            +{\rm S}(x) & \ruleeq & {\rm V_1}(x), \; \neg {\rm S}(x), \; 4 < x. \\
            -{\rm S}(x) & \ruleeq & \neg {\rm V_1}(x), \; {\rm S}(x), \; 4 < x.
        \end{array}
        $$
    \end{itemize}

    \item[Method]
    \begin{itemize}
        \item[]
        \item step 1: Derive well-behaved $get$ by inputtin {putdelta} into the putback-based BX.
        $$
        \begin{array}{lrl}
            {\rm V_1}(x)  & \ruleeq & {\rm S}(x), \; 4 < x.
        \end{array}
        $$

        \item step 2: Deform to $putdelta'$ by substituting definition of ${\rm V}(x)$ ((1) and (2)) into $putdelta$.
         $$
        \begin{array}{lrl}
            +{\rm S}(x)  & \ruleeq & {\rm V^{cur}_1}(x), \; \neg -V(x), \; \neg {\rm S}(x), \; 4 < x. \\
            +{\rm S}(x)  & \ruleeq & +{\rm V_1}(x), \; \neg {\rm S}(x), \; 4 < x. \\
            -{\rm S}(x)  & \ruleeq & \neg {\rm V^{cur}_1}(x), \; \neg +{\rm V_1}(x), \; {\rm S}(x), \; 4 < x. \\
            -{\rm S}(x)  & \ruleeq & -{\rm V_1}(x), \; \neg +{\rm V_1}(x), \; {\rm S}(x), \; 4 < x. \\
            {\rm V^{cur}_1}(x)    & \ruleeq & {\rm S}(x), \; 4 < x.
        \end{array}
        $$

        \item step 3: Derive $unde\!f$ by substituting $putdelta'$ into the definition of $unde\!f$ ((3) - (6)) and utilizing rules which contain equation or non-equation against $+{\rm V_1}(x)$ or $-{\rm V_1}(x)$.
        $$
        \begin{array}{lrl}
            +{\rm V_1^{ud}}(x)  & \ruleeq & +{\rm V_1}(x), \; \neg (4 < x). \\
            -{\rm V_1^{ud}}(x)  & \ruleeq & -{\rm V_1}(x), \; \neg (4 < x).
        \end{array}
        $$

        \item step 4: Derive well-behaved $get'$ by inputting $putdelta$ and $unde\!f$ into the putback-based BX.
        $$
        \begin{array}{lrl}
            {\rm V_1}(x)  & \ruleeq & {\rm S}(x), \; 4 < x. \\
            {\rm V_1}(x)  & \ruleeq & {\rm V_1^{ud}}(x), \; \neg (4 < x).
        \end{array}
        $$
    \end{itemize}

  \item[Output]
    \begin{itemize}
        \item []
        \item  Source: ${\rm S}(x), \; {\rm V_1^{ud}}(x)$
        \item  View:   ${\rm V_1}(x)$
        \item  $unde\!f$
        $$
        \begin{array}{lrl}
            +{\rm V_1^{ud}}(x)  & \ruleeq & \neg {\rm V_1^{ud}}(x), \; {\rm V_1}(x), \; \neg (4 < x). \\
            -{\rm V_1^{ud}}(x)  & \ruleeq & {\rm V_1^{ud}}(x), \; \neg {\rm V_1}(x), \; \neg (4 < x).
        \end{array}
        $$
        \item  $get'$
        $$
        \begin{array}{lrl}
            {\rm V_1}(x)  & \ruleeq & {\rm S}(x), \; 4 < x. \\
            {\rm V_1}(x)  & \ruleeq & {\rm V_1^{ud}}(x), \; \neg (4 < x).
        \end{array}
        $$
    \end{itemize}

 \end{description}

\vspace{\baselineskip}

Result shows a same behavior with Example 2.1.
The putback-based BX \cite{Van-dang1} translates derived rules of transformations into SQL that $get'$ as view definition and $putdelta$ and $unde\!f$ as trigger on view.
Then co-existing schemas are archived on relational database.

\section{Conclusion}

We presented the approach to achieve co-existing schemas by applying putback-based BX and introducing an extra function to satisfy bidirectionality and totality.
This provides flexibility to a user that a user arbitrary write behaviour of backward transformation for co-existing schemas without predefined restricted behaviour by others.
We contributed more flexibility to a user by systematically deriving proper transformation rules to achieve co-existing schemas.

As for the next steps, more complicated unsynchronized cases are to be studied.
Even for selection, it becomes complicated if two views share a same value from source.
Update of either one could be synchronized or unsynchronized.
Cases of join, projection and so on are to be studied.



\end{document}